\documentclass[pra,onecolumn]{revtex4}

\begin{document}

\title{Comment on `On quantum effects in spontaneous emission by a relativistic electron beam in an undulator' by G. Geloni, V. Kocharyan and E. Saldin (arXiv:1202.0691)}
%\shorttitle{Coherent and spontaneous emission in QFEL}
\author{G.R.M. Robb \& R. Bonifacio} 
\affiliation{SUPA, Department of Physics, University of Strathclyde,
Glasgow, G4 0NG, Scotland, UK }
%\ead{g.r.m.robb@strath.ac.uk}

\begin{abstract}
In arXiv:1202.0691, Geloni et al. criticise our recent work describing the spontaneous emission by a relativistic, undulating electron beam. In particular they claim that our prediction of a quantum regime in which evolution of the electron momentum distribution occurs as a sequential population of discrete momentum groups rather than in terms of a drift-diffusion process is unphysical and artificial. We show that the criticism of our model contained in arXiv:1202.0691 is unfounded
.
\end{abstract}
%\pacs{41.60.Cr,41.60.Ap,42.55.Vc}

\maketitle

\section{Introduction}
In \cite{Geloni}, Geloni et. al. criticised a recently published article \cite{spontaneous} in which we used a one-dimentional model to describe spontaneous emission by a relativistic electron beam moving through a magnetostatic undulator (or an electromagnetic pump field) as Compton backscattering.
Specifically they claim that the quantum regime described in \cite{spontaneous}, in which the electron momentum distribution evolves via sequential population of discrete momentum or energy groups, is artificial and unphysical. Geloni et al. claim that the reason for this so-called artificial result is the use of a one-dimensional model in \cite{spontaneous} rather than a more detailed three-dimensional treatment of spontaneous emission. They claim that the electron momentum or energy distribution should evolve as a continuous diffusive process completely described using a drift-diffusion equation. 

In the following we demonstrate why the criticism of the work in \cite{spontaneous} is unfounded.

\section{1-D versus 3-D model of spontaneous emission} 
While our use of a one-dimensional model to describe spontaneous emission is an approximation, it is an approximation which should be valid in the limit of a highly relativistic beam as considered in \cite{spontaneous} because the emission from the electron is concentrated in a narrow cone of solid angle $< 1 /\gamma^2 $ along the propagation direction of the electron beam, where $\gamma \gg 1$ is the relativistic factor \cite{Jackson}. This basic fact has been overlooked in \cite{Geloni}.

As part of the criticism of our 1-D model Geloni et al. remark that the electron is a point particle and not a plane charge sheet. The point-like particle is the classical picture of a single electron. The model in \cite{spontaneous} describes an ensemble of non-interacting electrons with a well-defined momentum in the axial direction. In quantum mechanics this ensemble is described by a wavefunction which is a plane wave.

\section{Quantum Regime Condition $\epsilon > 1$}
In \cite{spontaneous} the different regimes of emission were characterised using the parameter 
\begin{equation}
\label{epsilon}
\epsilon = \frac{\hbar k}{\sigma_p}
\end{equation}
, where $\sigma_p$ was defined in \cite{spontaneous} as a `characteristic momentum spread of the system' and $\hbar k$ is the Compton momentum recoil due to emission of a single photon. $\sigma_p$ is used to describe momentum spreads associated effects which are not explicitly included in the model. Consequently $\epsilon$ is not unphysical, as is claimed in \cite{Geloni}, but determines the significance of the discreteness of momentum recoil due to photon emission.
Geloni et al. state that the condition for the quantum regime stated in \cite{spontaneous} i.e. $\epsilon > 1$,  
is incorrect because the bandwidth of the spontaneous emission does not depend on the number of wiggler periods $N_w$, when $N_w \gg 1$ and that consequently our analysis is invalid. Here we demonstrate that this is not the case. Note that the definition of $\epsilon$ as stated in \cite{Geloni} is not the general one as originally defined in \cite{spontaneous} where two possible choices for $\sigma_p$ were considered, only the first of which is mentioned in \cite{Geloni} :
\begin{enumerate}
\item that associated with the finiteness of the interaction length, for which  $\sigma_p \approx \frac{\gamma m c}{N_w}$
\item that associated with high-gain FEL amplification, for which $\sigma_p \approx \rho \gamma m c$
\end{enumerate}

Geloni et al. criticise the fact that in the first of these cases $\sigma_p \propto 1/N_w$, arguing that this is an artifact of our 1-D model. It should be noted that the $1/N_w$ dependence is a consequence of the finite duration of the interaction length, which is not affected by the direction of emission i.e. no matter what direction the radiation is emitted it consists of a finite wavetrain of duration $N_w$ periods. This is in accordance with \cite{Jackson} and eq.(5) of \cite{Geloni}, which shows that the linewith of emitted radiation $\propto 1/N_w$ for any direction, not just on axis. However more important is the fact that the choice of $\sigma_p$ does not affect the basic assumptions of our model, which are summarised by equations (1) and (2) of \cite{spontaneous}, and are that the spontaneous emission process for a relativistic ensemble of electrons is well described by a 1-D model in which electron emit photons randomly in time according to a Poisson distribution as described in \cite{spontaneous}. 

\section{Modelling Spontaneous Emission Using a Drift-Diffusion Equation}
Geloni et al {\em assume} a drift-diffusion equation (eq.(7) of \cite{Geloni}) which they state is valid as long as the condition 
\begin{equation}
\label{zeta}
\zeta = \frac{\hbar \omega}{\gamma m c^2} \ll 1
\end{equation}
where $\omega = c k$. In constrast we {\em derive} a drift diffusion equation as a limiting case of our finite-difference equation when $\epsilon < 1$, where $\epsilon$ is defined in eq.~\ref{epsilon}. We stress that if we choose $\epsilon = \frac{\hbar \omega N_w}{\gamma m c ^2} < 1$ then the drift-diffusion equation assumed in \cite{Geloni}(eq.(11) in \cite{Geloni}) and derived in \cite{spontaneous} (eq.(7) in \cite{spontaneous}) have the same coefficients to within a numerical factor, which is acknowledged in \cite{Geloni}. Note that eq.~(\ref{epsilon}) can be written as 
\begin{equation}
\label{epsilon2}
\epsilon = \frac{\hbar \omega}{\sigma_p m c^2} =\frac{\zeta}{(\sigma_\gamma / \gamma)}
\end{equation}
where $\sigma_\gamma$ is the spread in $\gamma$ for the electron ensemble. Consequently there is no conflict between the condition in eq.(~\ref{zeta}) and $\epsilon > 1$ as long as $\frac{\sigma_\gamma}{\gamma} \ll 1$, which is true for all systems of practical interest. The condition in eq.~(\ref{zeta}) means that it is possible to neglect the Compton wavelength shift during the emission process and that the recoil momentum due to photon emission can be neglected when compared to the electron momentum. However the condition $\epsilon < 1$ means that the single-photon momentum recoil is less than the {\em spread} in electron momenta, $\sigma_p$. The analysis of Geloni et al. is restricted to the case of a single electron so the electron momentum spread is not a quantity which can be used for normalisation. The model of \cite{spontaneous} in contrast describes an ensemble of electrons, which has a finite spread in momenta. The condition $\epsilon > 1$ therefore describes the case where photon recoil momentum is larger than the electron momentum spread and gives rise to discrete momentum groups so that a description in terms of a continuous drift-diffusion equation is no longer valid, even if $\zeta \ll 1$, contrary to what is claimed in \cite{Geloni}. 
  
\section{Conclusion}
In conclusion, we have demonstrated that the criticism of our model \cite{spontaneous} by Geloni et al. in \cite{Geloni} is unfounded because it considers only a single electron rather than an ensemble with some momentum spread, and that there is no conflict between the condition for the appearance of discrete electron momentum groups ($\epsilon > 1$) and the condition $\zeta \ll 1$ which allows the Compton wavelength shift to be neglected.

\end{document}